\documentclass[a4paper, 
10pt
]{article}

\usepackage[a4paper,left=3cm,right=3cm,top=2.5cm,bottom=2.5cm]{geometry}
\usepackage{amsmath}
\usepackage{multirow}
\usepackage[table]{xcolor}

\usepackage{subcaption}
\usepackage{pgfplots}
\usepackage{pgfplotstable}
\usepgfplotslibrary{groupplots}
\usepgfplotslibrary{fillbetween}
\usepackage{mathtools}
\usepackage{multicol}
\usepackage{comment}
\usepackage{booktabs}
\pgfplotsset{compat=1.5}
\usepackage{amssymb}
\usepackage{url}
\usepackage{bm}

\usepackage{float}
\usepackage{tabularx}
\usepackage{enumerate}
\usepackage{array}
\usepackage{xspace}
\usepackage{tikz}
\usepackage{tikzsymbols}
\usetikzlibrary{calc,trees,positioning,arrows,chains,shapes.geometric,%
    decorations.pathreplacing,decorations.pathmorphing,shapes,%
    matrix,shapes.symbols, decorations.markings, patterns,fit,quotes,%
    arrows.meta}

\usepackage{authblk}
\usepackage{xcolor}
\usepackage{colortbl}

\definecolor{Gray}{gray}{0.9}

\usepackage{hyperref}
\usepackage{siunitx}

\usepackage[bottom]{footmisc}

\newcommand{\switch}{$SWITCH$\xspace}
\newcommand{\user}{$USER$\xspace}
\newcommand{\hub}{$HUB$\xspace}
\newcommand{\source}{$SOURCE$\xspace}

\newcommand{\edgeor}{$OR$\xspace}
\newcommand{\edgeand}{$AND$\xspace}
\newcommand{\edgesingle}{$SINGLE$\xspace}

\newsavebox{\largestimage}

\begin{document}

\date{}
\title{A graph-based framework for complex system simulating and diagnosis with automatic reconfiguration} 

\author[]{Martina Teruzzi\footnote{martina.teruzzi@sissa.it}}
\author[]{Nicola Demo\footnote{nicola.demo@sissa.it}}
\author[]{Gianluigi Rozza\footnote{gianluigi.rozza@sissa.it}}

\affil{Mathematics Area, mathLab, SISSA, via Bonomea 265, I-34136
  Trieste, Italy}

\maketitle

\begin{abstract}
Fault detection has a long tradition: the necessity to provide the most accurate diagnosis possible for a process plant criticality is somehow intrinsic in its functioning. Continuous monitoring is a possible way for early detection. However, it is somehow fundamental to be able to actually simulate failures. Reproducing the issues remotely allows to quantify in advance their consequences, causing literally no real damage. Within this context, signed directed graphs have played an essential role within the years, managing to model with a relatively simple theory diverse elements of an industrial network, as well as the logic relations between them.\\
In this work we present a quantitative approach, employing directed graphs to the simulation and automatic reconfiguration of a fault in a network. To model the typical operation of industrial plants, we propose several additions with respect to the standard graphs: {\it i}) a quantitative measure to control the overall residual capacity, {\it ii}) nodes of different categories --- and then different behaviors --- and {\it iii}) a fault propagation procedure based on the predecessors and the redundancy of the system. The obtained graph is able to mimic the behaviour of the real target plant when one or more faults occur. Additionally, we also implement a generative approach capable to activate a particular category of nodes in order to contain the issue propagation, equipping the network with the capability of reconfigure itself and resulting then in a mathematical tool useful not only for simulating and monitoring, but also to design and optimize complex plants.
The final asset of the system is provided in output with its complete diagnostics, and a detailed description of the steps that have been carried out to obtain the final realization.
\end{abstract}


\section{Introduction}
\label{sec:intro}
Fault detection is fundamental for industrial equipment: it allows to detect potentially destructive situations, preserving safety and functioning of the system at best. For this reason, a lot of efforts have been spent in its modeling and diagnostics \cite{fault-spanners, vertex_sets, detectors}. Graphs have met the necessity to model precisely process plants, leading the path to the creation of ad-hoc softwares \cite{multilevel_flow_modeling, equipment_model_builder}. Moreover, the qualitative paradigm of signed directed graphs has been enriched within the years with high resolution techniques, able to provide a quantitative description \cite{sdg_qta, mpsdg, sdg_nrs}.

Besides a precise description of fault consequences, the ability to simulate these kinds of fatal events creates the opportunity for real-time intervention, aimed at an automatic reconfiguration of the system with its largest residual operational capacity.\newline
In this work we are going to present a screening tool we have developed within this framework, able to mimic 
different kinds of perturbations in interconnected systems, such as industrial plants. The dependencies between the network elements is indeed depicted employing graph theory, such that the results obtained from the graph analysis can therefore be used to improve the robustness and resilience profile of industrial facilities against domino effect propagation.
Not only: we coupled the graph model with a custom fault propagation procedure --- based on logical relations between the components --- to provide a simplified \emph{digital twin} of the studied plant that behaves like the physical one. This finally allowed us to employ a genetic procedure that, controlling the active propagation resistances, is able to dynamically change such resistances states in order to limit the effect of a generic fault.\newline
All the features described in the present work have been implemented in SAFEX software, within project SAFE,  “Realtime Damage Manager And Decision Support”. An open-source restricted version of the software can be found in GRAPE (GRAph Parallel Environment) software package \cite{GRAPE, AuroraMHPC, TeruzziMHPC}.

The paper is organized as follows: in Section 2, after an initial digression on graph definitions that are going to be valid throughout the whole dissertation, two algorithms for the calculation of shortest-paths are presented. After that, the measure of service is introduced, together with the benefits of its adoption. Moreover, a fault propagation simulation example is described on a toy graph, with a final explanation of the generative approach employed for SWITCH nodes activation. Section 3 shows more realistic applications: the first one concerns a switch line, extremely useful for an exhaustive illustration of how active resistance to perturbations is carried out. The second one, instead, is a characterization of the results for simulations on random graphs.

\section{Other section}
\label{sec:other}
Graph theory is a well-established mathematical framework. Its origins date back to Euler in 1736 \cite{Euler}; from the very beginning its rigorous formalization has been entangled with scientific and technological development in various fields, including mathematics~\cite{Sylvester, ErdosRenyi}, chemistry~\cite{Cayley}, physics and engineering~\cite{Kirchhoff}.
This section is devoted to introduce the methodology we have applied for the modeling of complex systems: initially we provide a formal overview about graph theory, defining the main features of a generic graph, to then focus about the introduced characteristics and the genetic calibration of the graph itself.

\subsection{Graph definitions}
A graph can be defined as a general representation of a structure describing relations between objects. For its generality, it is well-suited for the description of very diverse structures at many different scales \cite{graphs_circuits, graphs_bio, graphs_socialnetwork}. In the current application, graphs have been employed in the description of interconnected systems, such as industrial plants.\newline
We define a graph \textit{G} as formed by a set of nodes \textit{N} and a set of edges \textit{E}; each edge connects two nodes. If a \textit{weight} is assigned to each edge (usually a positive integer value), the graph is called \textit{weighted}; \textit{unweighted}, otherwise. A directed graph (or \textit{digraph}) is a graph whose edges have a direction, with one node as head and the other as tail. The graph is called \textit{sparse} if $|E|\propto |N|$, while it is called dense if $|E| \propto {|N|}^2$. The length of a \textit{walk} (intended as a succession of alternating nodes and edges) between two nodes \textit{u} and \textit{v} is given by the number of the edges traversed, if the graph is unweighted. Otherwise, the length is given by the sum of the edge weights. A node \textit{v} is said to be reachable from a node \textit{u} if there exists at least one walk from \textit{u} to \textit{v}. A walk with no repeated nodes is called a \textit{path}. The \textit{distance} between two nodes \textit{u} and \textit{v} is the length of the shortest walk containing them. Within the current application, we are always going to address weighted directed graph.
\subsection{The system graphs}
The proposed methodology aims to model a generic plant through a graph: the components that compose such plant --- e.g. a water pump, a valve --- are indeed modeled as nodes, while the connections between such objects --- the reader can think of cables and wires --- are simulated through the graph edges.
In order to provide a more realistic description of how commodities flow within an industrial plant, the graphs have been equipped with weighted edges. Weight can be very useful in specifying preferable itineraries for service stream, with respect to areas in which its course can be less favored. This feature is of immediate integration within the description of the network with graphs, allowing a very efficient calculation of the shortest path as the path with the minimum length.

\begin{figure}
  \centering
  \includegraphics[width=0.4\textwidth]{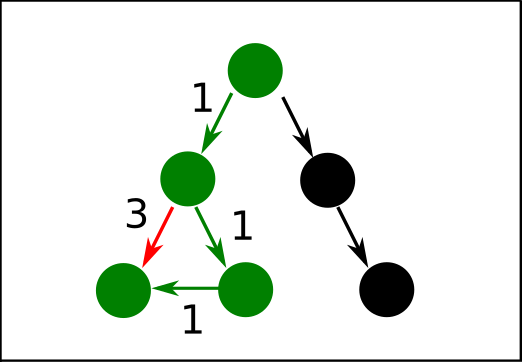}
  \caption{Shortest Path. In this weighted digraph $G(N,E)$ the shortest path is indicated in green, with respect to the red path, which implies a larger weighted distance between the nodes.}
  \label{fig:shortest_path}
\end{figure}

A shortest path is defined as the walk between two vertices that minimizes the sum of the weights of the edges traversed. An example is given in Figure~\ref{fig:shortest_path}. Shortest-path calculation plays a crucial role in graph analysis. It is necessary in order to be able to compute a series of quantities, known as efficiency and centrality measures, for system diagnostics.
For weighted graphs we have implemented two algorithms for shortest-path calculation: Dijkstra's method \cite{Dijkstra} for sparse graphs and Floyd-Warshall \cite{Floyd, Roy, Warshall} for dense graphs.\newline
Dijkstra’s algorithm assigns some initial distance values, improving them step by step. It starts with marking all the nodes as unvisited, assigning to every node a tentative distance value, which is set to zero for the initial node (that will be
labeled as current), while infinity for all other nodes. The search proceeds between all the unvisited neighbors of current node, calculating their tentative distance through it, and updating if smaller than the previous one. When all its neighbors have been visited, current node is marked as visited and never checked again. For all-pairs distance this algorithm is of order $O(|N|(|E| +|N|log(|N|)))$.\newline
Floyd-Warshall algorithm, instead, involves the comparison of all the possible routes between a pair of vertices \textit{i} and \textit{j}. If the distance is smaller than the previous one, distance and predecessor matrices get updated. A final path reconstruction returns the shortest paths for the graph.

From shortest paths it is possible to extract the path \textit{efficiency}, which is given by the inverse of its length, when different from zero, and zero otherwise. We recall that, for weighted graphs, the path length does not depend on the number of edges traversed, but on the sum of their weights. From this local measure, further measures can be obtained, providing a more global view. Our implementation also provides \textit{centrality} measures. These quantities are more related to how well-connected a node is, and their comparison before and after a fault can be a useful hint on how the situation has changed due to the perturbation. Shortest-path and measures calculations are repeated after the propagation of a damage, and their description for the integer and damaged graph represents an important part of the system diagnosis. More details about efficiencies and centralities can be found in \cite{GRAPE}.

\subsection{Graph residual service}
\begin{figure}
  \centering
  \includegraphics[width=\textwidth]{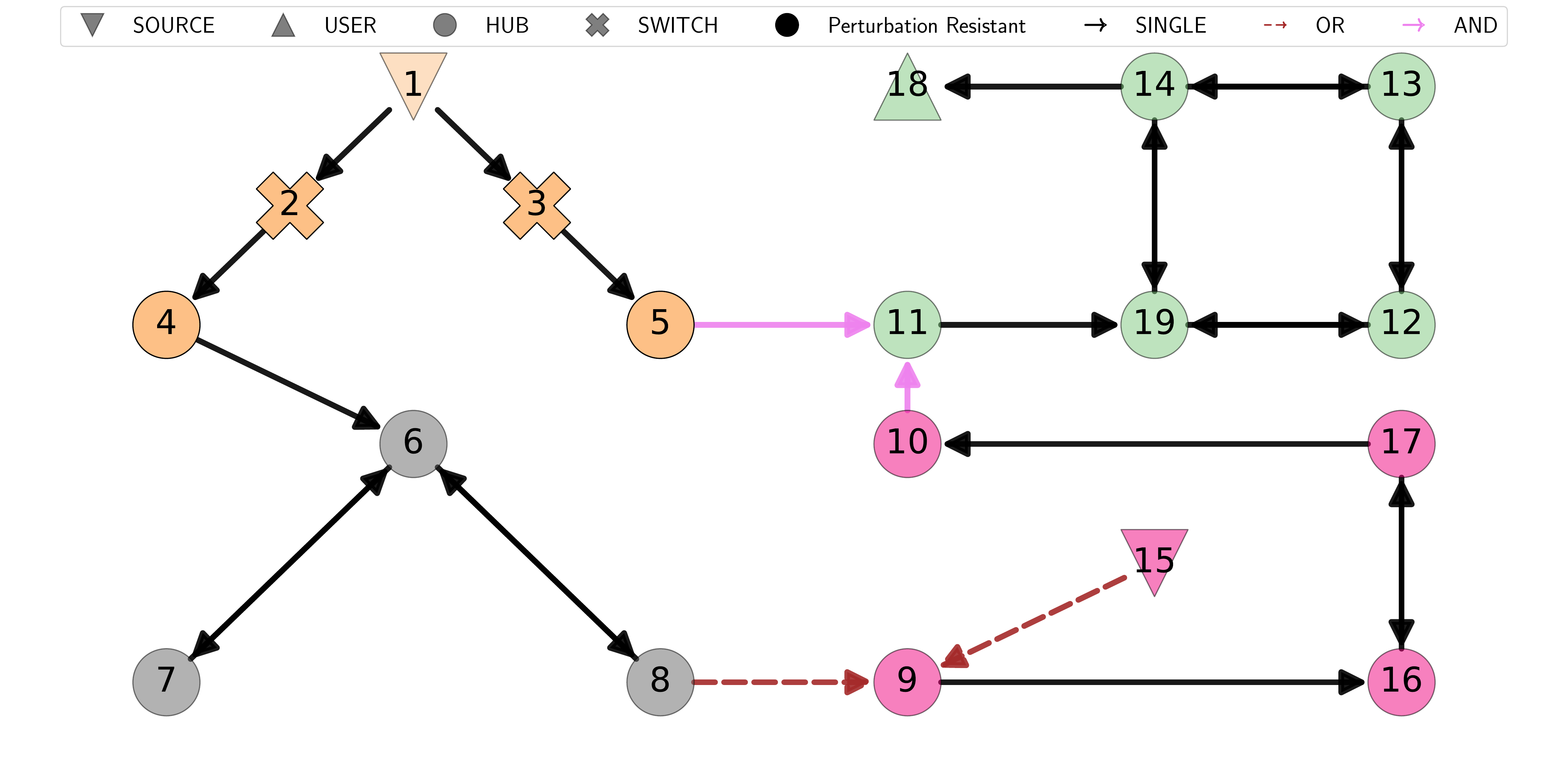}
  \caption{Example graph. The four different color identify different areas. Nodes are labeled as \source (reverse triangle), \hub (circle) or \user (up triangle), while the edges can describe \edgeor, \edgesingle, or \edgeand relations between linked nodes. \switch nodes (cross) are a particular type of \hub nodes. Nodes with no transparency are passive resistant nodes.}
  \label{fig:TOY_graph}
\end{figure}

Meaning a graph as a collection of elements belonging to a system in which commodities flow, like an industrial plant, it becomes fundamental to be able to measure the state of this flow, especially after the advent of a perturbation.
In addition to the aforementioned measures, we then introduce a feature representing the quantitative description of the system residual capacity. We refer to this measure for the rest of the article as residual service, or just \textit{service}. To our knowledge, no similar property has been implemented before for the flow of commodities.\\
The importance of service, together with its intrinsic relation with fault propagation, has an immediate effect on the graph characterization. In Figure~\ref{fig:TOY_graph} a simple example graph is reported; despite its conciseness, it contains all the key ingredients that are necessary to comprehend the whole picture.
Nodes have been labeled regarding their role when service is taken into account: \source nodes provide service to the network, while \user nodes deplete it. \hub nodes represent the rest, and they usually constitute the paths going from \source to \user. It is important to highlight that \source and \user nodes can be seen not just as components, but also as the input and output coming from other sub-systems. In this way we can connect different sub-structures, which exchange information using service. It is also important to specify that for each \source, service has to be split between all survived \user nodes connected to it. If some \user nodes are broken due to a perturbation, the residual service is re-computed taking just the available ones into account.

\subsection{Fault propagation}
In addition to definition of service, we add the possibility to specify three different logical relations between nodes connected by an edge (Figure~\ref{fig:TOY_graph}), allowing for modeling a larger variety of possible systems.
\edgesingle edge connects a node to its unique predecessor. \edgeand edge reports the fact that a node has more than one predecessor: all the predecessors are necessary for the functioning of that component. \edgeor edge also reports the fact that a node has more than one predecessor: however, in this case just one of the predecessors should be active to guarantee the functioning of that component. Finally, if a node has no predecessor it is labeled in the input as $ORPHAN$.\\
Regarding the behaviour of the graph dealing with a generic fault, we introduce two kinds of perturbation resistances that can confine the cascade effects, which are highlighted in Figure~\ref{fig:TOY_graph}. The first one is passive resistance, for nodes that are not affected by the perturbation. This kind of resistance is intrinsic of the component itself (one example can be a fire-proof element within a plant). The second resistance is called active, implying the voluntary activation of some graph elements in the damage propagation process, preventing the perturbation involving neighboring components. This particular nodes that can be turned on or off in order to block the cascade are called \switch nodes. By convention in SAFEX we set a \switch to \textit{True} if it allows the flow of commodities (and, in this way, also the propagation of damage), like in a closed circuit. Vice-versa, a \switch is set to \textit{False}.

\subsection{A fault simulation example}
\begin{figure}
  \centering
  \includegraphics[width=0.9\textwidth]{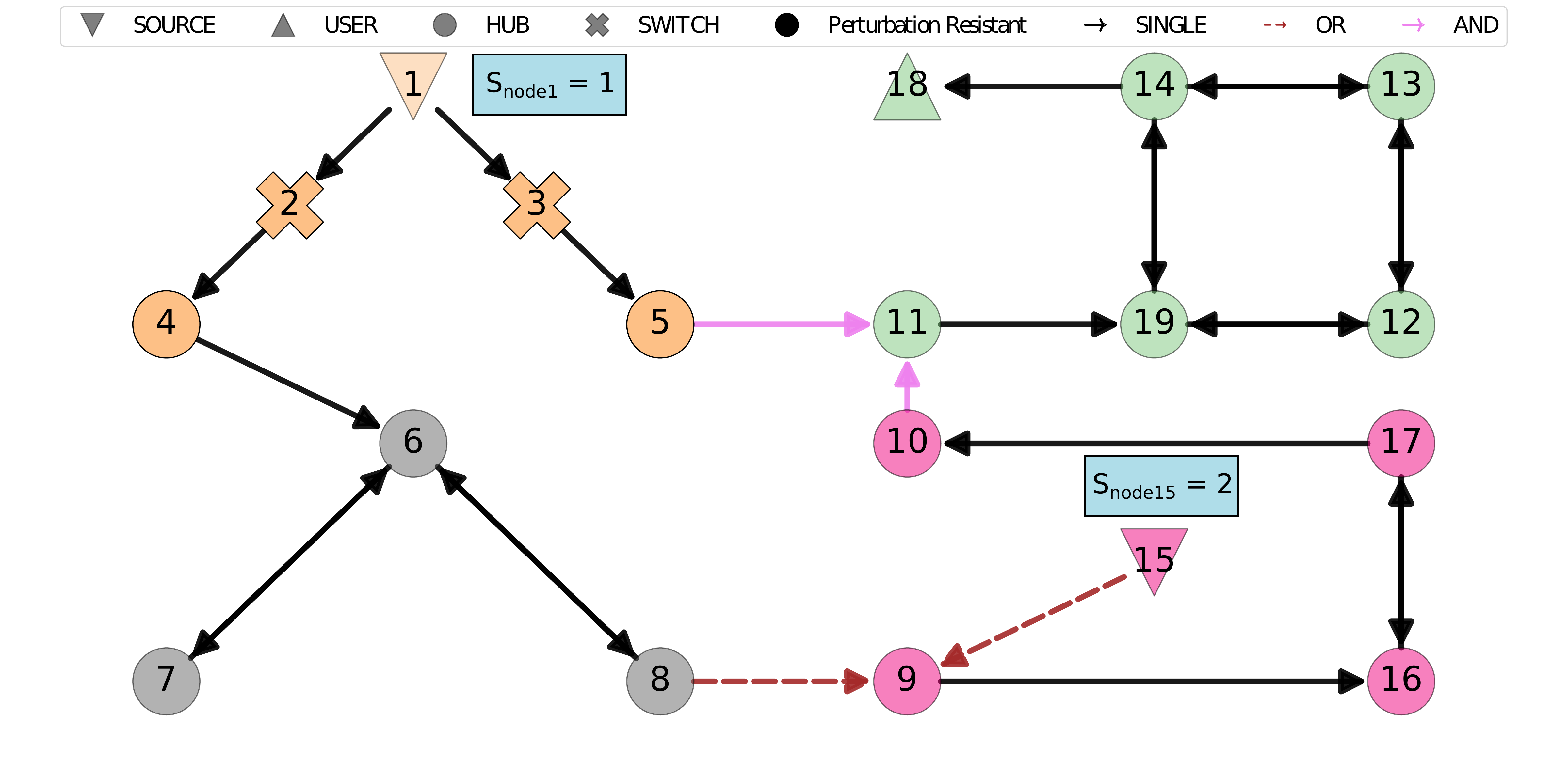}
  \includegraphics[width=0.9\textwidth]{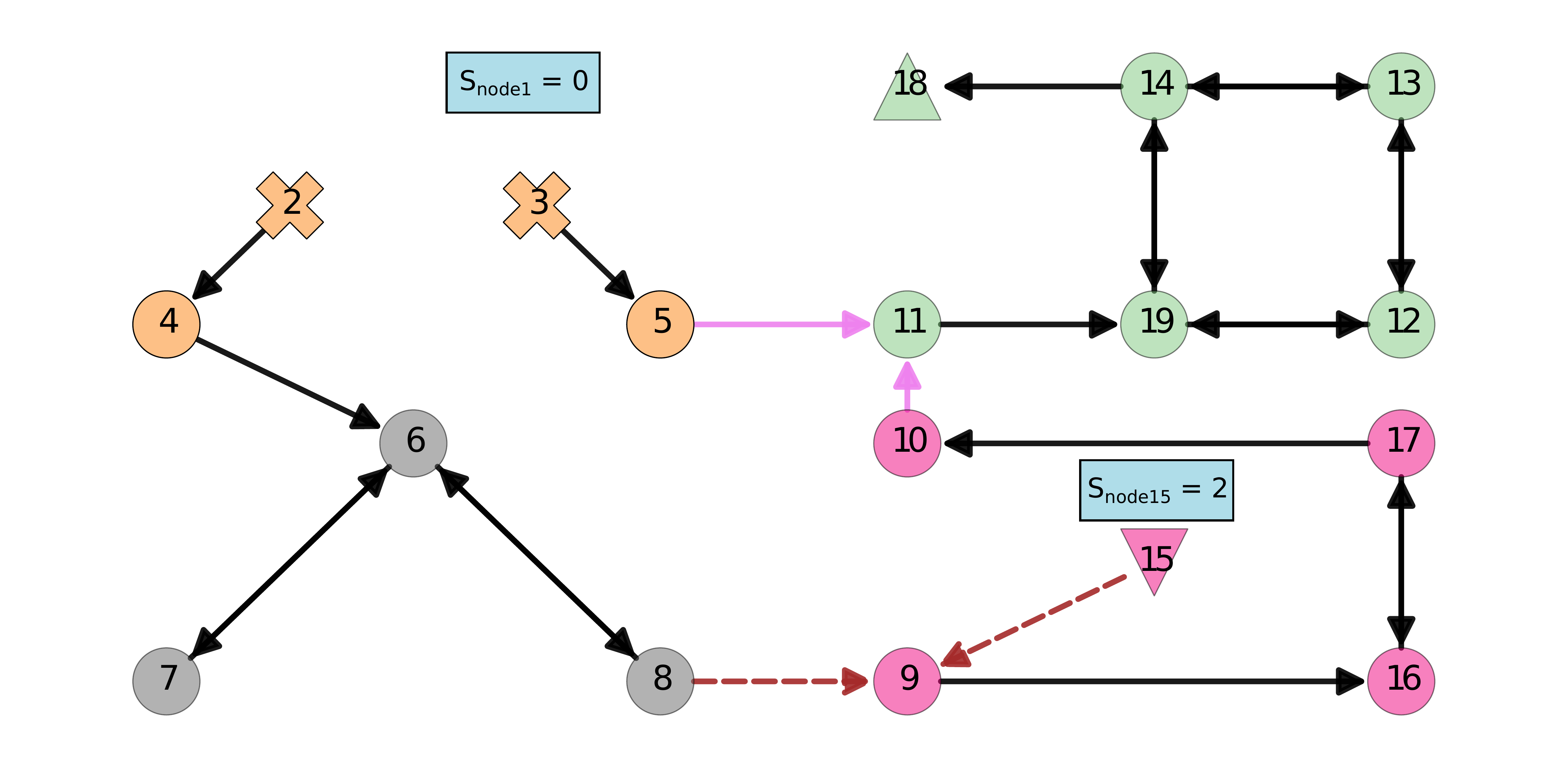}
  \caption{Example of a fault event, the damage of one node, namely \source $1$. On top, the integer graph, while the damaged one at bottom. $S_{node1}$ and $S_{node15}$ represent the service provided by nodes $1$ and $15$, before and after the perturbation.}
  \label{fig:simulate_perturbation}
\end{figure}

After having explained all the main characteristics of our graphs, in this section we are going to present one simple fault event, in order to give a picture of the propagation procedure. In general, perturbations can affect more than one node at the same time.\newline
In SAFEX you can perturb the graph in different ways; we are going here for simplicity to consider the perturbation of one node, namely node $1$. In Figure~\ref{fig:simulate_perturbation}, on the left, the integer graph is shown. There are two \source nodes, namely $1$ and $15$, both connected to the only \user, node $18$. They provide service $S_{node1} = 1$ and $S_{node15} = 2$, respectively, so that the total service at the user $18$ is $3$. When node $1$ is perturbed, not possessing any resistance to the damage, it gets broken. \switch nodes $2$ and $3$ are opened (e.g., set to \textit{False}), so that the cascade effect is limited to node $1$ only. We are going to discuss the way \switch nodes get activated further in this paper. For now, we conclude that the simulation results just in the loss of node $1$, while the total final service at node $18$ is $2$, since \source node $1$ is broken and no more able to provide service.
\subsection{Switch activation: a generative approach}
\label{subsec:genetic}
\begin{figure}[htbp]
  \centering
  \savebox{\largestimage}{\includegraphics[height=.15\textheight]{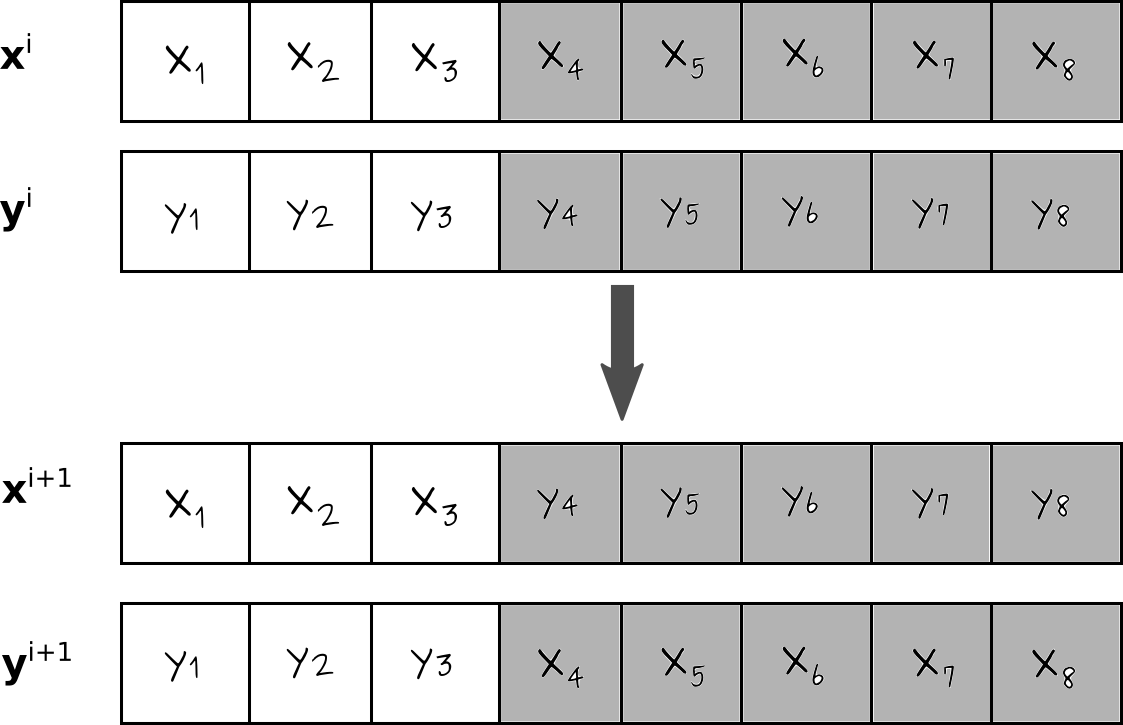}}%
  \begin{subfigure}[b]{0.48\textwidth}
    \centering
    \usebox{\largestimage}
    \caption{Crossover.}
    \label{fig:crossover}
  \end{subfigure}
  \quad
  \begin{subfigure}[b]{0.48\textwidth}
    \centering
    \raisebox{\dimexpr.5\ht\largestimage-.5\height}{%
      \includegraphics[height=.08\textheight]{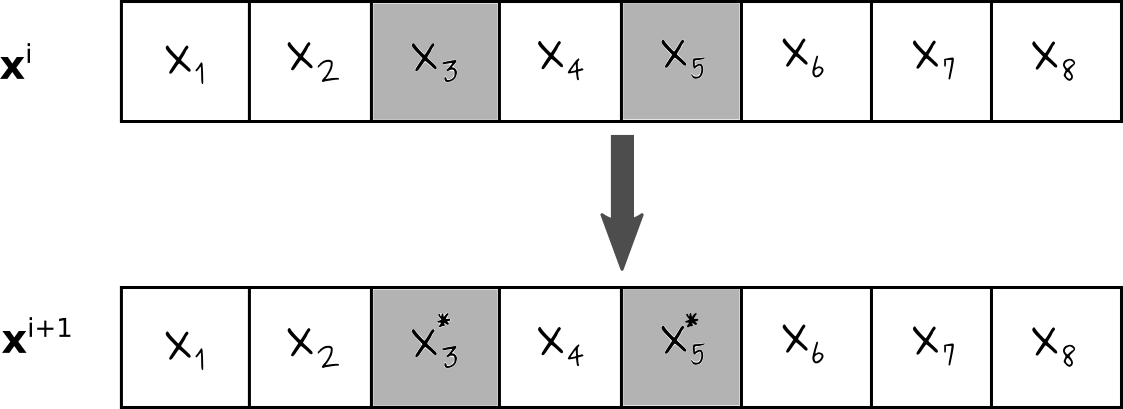}}
    \caption{Mutation.}
    \label{fig:mutation}
  \end{subfigure}
  \caption{Graphical sketch of crossover and mutation procedures.}
\end{figure}

The example graph in Figure~\ref{fig:TOY_graph} shows just two nodes labeled as \switch. For this simple case, we can list all the possible configurations that they can assume very easily. Increasing the number of \switch nodes, the number of possible realizations increase exponentially. Therefore, determining the best configuration for a certain simulated fault is not straightforward.\newline
In order to solve this problem in general, in SAFEX we employ a generative approach. Genetic algorithms date back to 1973 \cite{HollandI, HollandII}, and they have had a strong development within the years \cite{genetic_review, genetic_hybrid, genetic_selection, genetic_optimization, genetic_quadratic}, still maintaining their basic concept, which we are going to summarize here to help a better understanding.\newline
These kinds of algorithms generate a population of individuals with random genes, and make them evolve mimicking Darwin's theory of evolution. The main steps are \textit{selection}, \textit{crossover} and \textit{mutation}.\newline
We can define a population as composed by $N$ individuals ${\mathbf{x}}_i \in {\mathbb{R}}^P$ with $P$ genes as $\mathbf{X} = \{{\mathbf{x}}_i,...,{\mathbf{x}}_N\}$. We express the fitnesses of each individual using a scalar function $f:{\mathbb{R}}^P\rightarrow\mathbb{R}$. The first generation ${\mathbf{X}}^1$ is generated randomly, and for every individual in the population the respective fitness is computed: $y_i = f({\mathbf{x}}_i)$ for $i=1,...,N$. Every iteration (called \textit{generation}), implies the following steps:
\begin{itemize}
    \item \textit{selection}: the best individuals of the previous generation are chosen, according to their fitnesses, in order to form the new one. The number of individuals chosen and the approach can vary;
    \item \textit{crossover}: the selected individuals are combined in pairs, with a certain probability. Mixing the genetic information of the two parents allows the creation of one or more new offsprings, which are going to constitute the new generation ${\mathbf{X}}^{i+1}$. You can see a graphical sketch of this procedure in Figure~\ref{fig:crossover};
    \item \textit{mutation}: one or more individuals evolve changing part of their genes, according to some probability. You can see a graphical sketch of this procedure in Figure~\ref{fig:mutation}.
\end{itemize}
The total number of generations is itself a parameter of the algorithm. Iterating over generations, it is possible to optimize the original population. In the actual implementation of the genetic algorithm in SAFEX for the optimization of \switch nodes, we have taken advantage of DEAP (Distributed Evolutionary Algorithms in Python) library \cite{DEAP}.\newline
We still need to define what is an individual, and how fitness is evaluated for it, in the current framework. A \textit{state} is defined as one of the possible realizations for \switch nodes: hence, ${\mathbf{x}}_i$ is a list of $True$ and $False$ values, with dimension equal to the total number of these nodes. The set of SWITCH nodes is defined in input. For each of the states, three quantities constitute the fitness:
\begin{itemize}
    \item $n_{actions}$: the total number of ``flips'' necessary in order to turn the initial configuration (given in input) into the one represented by the individual. The best configuration is supposed to be the closest to the initial with the lowest fitness. Within a real industrial plant, whether the activation of a \switch is manual or automatic, the quickest solution remains the one with the smallest intervention;
    \item $S_{tot}$: total final residual service, given by the sum of the residual service at all \user nodes;
    \item $n_{alive}$: total number of survived nodes after the perturbation, for the considered configuration of \switch nodes.
\end{itemize}
The weights for this three different quantities are set by the user; in any case, we want to minimize the first quantity, while maximizing the other two, so that the fitness reads:
\begin{equation}
    y_i = f({\mathbf{x}}_i) = w_{1}\cdot{n}_{actions}({\mathbf{x}}_i) - w_{2}\cdot{S}_{tot}({\mathbf{x}}_i) - w_{3}\cdot{n}_{alive}({\mathbf{x}}_i)
\end{equation}

\section{Numerical results}
\label{sec:results}
In this section we are going to show some numerical results we have obtained 
by applying the described modeling technique to different test cases. We remark again that, concerning the implementation side, our framework has been implemented in SAFEX, which of course has been employed also to collect the results we are going to discuss.
In the first part, we will focus on a demonstration of the activation of \switch nodes on a small graph, testing the introduced methodologies for simulating a perturbation. Thanks to the limited dimension of the plant (therefore of the corresponding graph) we are able to show in detail all the steps computed for propagating such perturbation, as well as a qualitative discussion about the outcome of the genetic re-calibration. In the second part we analyze larger graphs in order to quantitatively measure the effectiveness of the proposed framework.
\subsection{Switch line}
\begin{figure}
  \centering
  \includegraphics[width=1.0\textwidth]{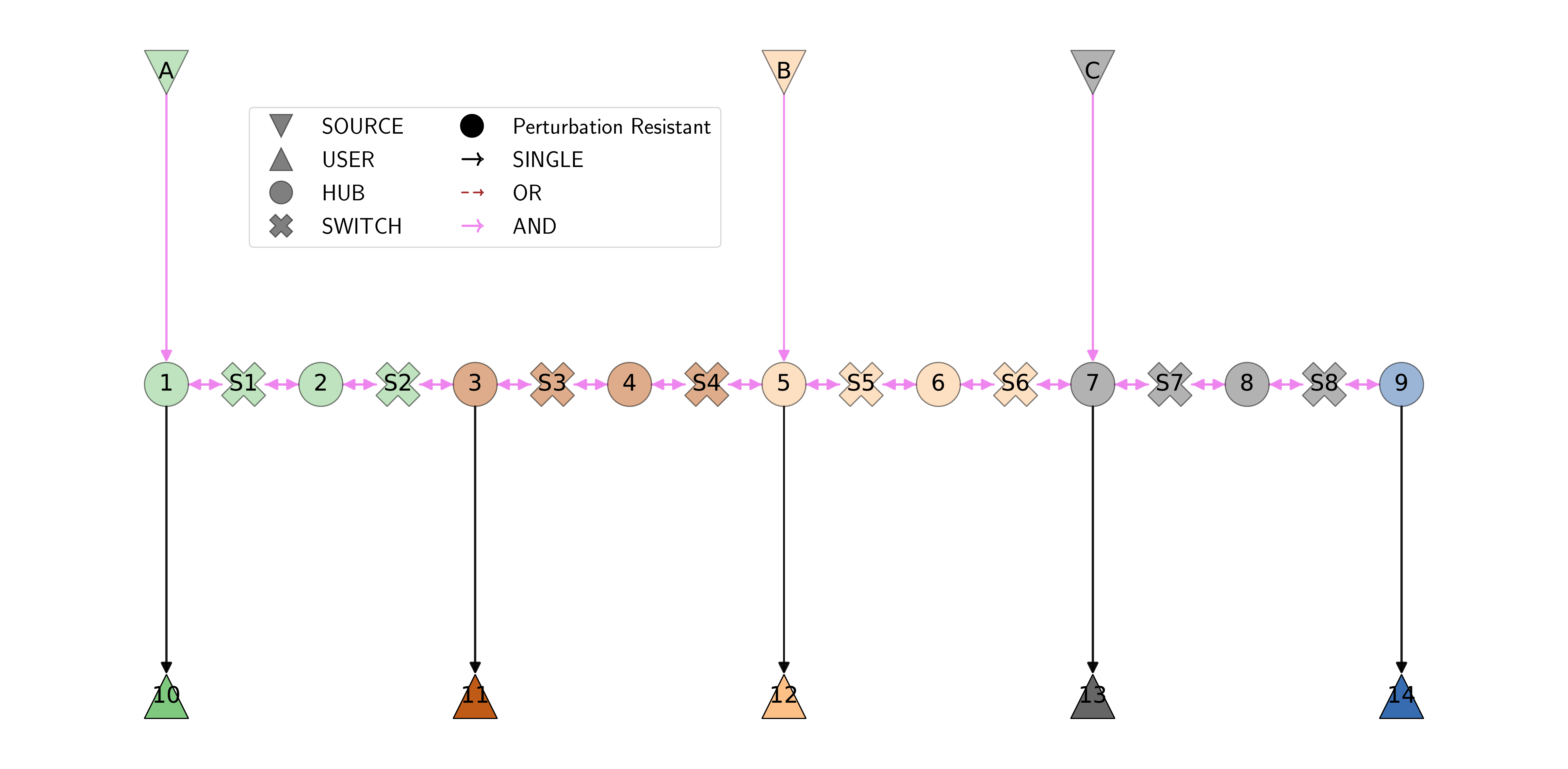}
  \caption{Switch line. This graph is divided in five areas, which are identified by the different color. Nodes are labeled as \source (down triangle), \hub (circle), \switch (cross) or \user (up triangle), while the edges can describe \edgeor, \edgesingle, or \edgeand relations between linked nodes. Nodes with no transparency are passive resistant nodes.}
  \label{fig:switch_line}
\end{figure}

The first system under study is a very good illustration of the mechanism for the activation of \switch nodes. The graph in Figure~\ref{fig:switch_line} shows, in fact, a sequence of nodes interposed with \switch nodes. The possible states are many: for this reason we can clearly see the genetic algorithm described in Section~\ref{subsec:genetic} in action.\newline
The graph in Figure~\ref{fig:switch_line} includes 25 elements connected by direct edges that reflect the hierarchy of the system in a parent-child fashion. 
The nodes are distributed in adjacent areas:
\begin{itemize}
    \item $area1$ includes 6 nodes, namely $A$, $1$, $S_1$, $2$, $S_2$, and $10$;
    \item $area2$ includes 5 nodes, namely $3$, $S_3$, $4$, $S_4$, and $11$;
    \item $area3$ includes 6 nodes, namely $B$, $5$, $S_5$, $6$, $S_6$ and $12$;
    \item $area4$ includes 6 nodes, namely $C$, $7$, $S_7$, $8$, $S_8$ and $13$;
    \item $area5$ includes 2 nodes, namely $9$ and $14$.
\end{itemize}
In the graph in Figure~\ref{fig:switch_line}, all edges are \edgeand edges. This choice has been made in order to have the widest possible spread of any perturbation. A perturbation of one or multiple elements in one area may exceed the area boundaries and propagate to other components connected to it, located in other areas. Nodes $10$, $11$, $12$, $13$, $14$ are perturbation resistant nodes, showing passive resistance. Hence, these nodes will not be affected by the simulated perturbation. Nodes $S_1$, $S_2$, $S_3$, $S_4$, $S_5$, $S_6$, $S_7$ and $S_8$ are \switch isolating elements.\newline
As stated previously, \switch nodes activation can stop the propagation of a fault in a graph, avoiding a very dangerous cascade effect. They are characterized by two possible values: \textit{True} or \textit{False}. For this example, we are going to set the genetic algorithm weights to unity ($w_1 = w_2 = w_3 = 1$). Moreover, we set the \switch nodes all to \textit{True} as initial condition:
\begin{equation}
    {\mathbf{x}}_{initial} = \{S_i =True, \quad \forall i =1,...,8 \}
\end{equation}
If all the three sources $A$, $B$ and $C$ in Figure~\ref{fig:switch_line} provide a service equal to $1$, then for the initial configuration chosen each of the user obtains an amount of service equal to $3/5 = 0.6$.\\
The first perturbation we would like to analyse is the perturbation of node $1$. Node $1$ has no fault resistance of any kind; hence, it gets broken.
After that, the perturbation could in principle propagate to all nodes (if not limited); however, the genetic evolution identifies the following best state: all the switches but $S_1$ remain $True$ (hence, closed). This configuration has the following characteristics in terms of fitness: $n_{actions} = 1$ with respect to the initial condition, total final service $S_{tot} = 2$ (\source node $A$ is no more connected to any of the \user nodes), and $n_{alive} = 24$ nodes surviving to the perturbation. In conclusion, $y_{best} = -25$ is the fitness of the best configuration. Notice that setting a \switch node to $False$ cuts off the edges between that node and its predecessors.\newline
The second result for the switch line graph concerns the perturbation of node $2$. This time, we perturb again a single node; however, this node is surrounded by more switches. For this reason, in order to isolate the broken node, we need to open two switches: both $S_1$ and $S_2$ get activated. The fitness of the best state can be computed like: $y_{best} = 2-3-24 = -25$. It is important to notice that opening $S_1$ and $S_2$ has indeed modified the flow in the graph. In fact, the final service is $S_{tot} = 3$, as for the initial configuration, but differently distributed between the \user nodes with respect to the starting point. While initially all the \user nodes were evenly receiving $1/5$ of the total service, in the final situation node $10$ is the only one still connected to \source $A$, receiving final service $1$, while the other four \user nodes have all final service equal to $0.5$, splitting evenly the service provided by $B$ and $C$.\newline
As a final result for the switch line, we would like to perturb more than one node, namely nodes $2$ and $3$. This kind of perturbation is interesting because, differently from the previous one inspected here, if we set weights $w_1 = w_2 = w_3 = 1$, then two states have the same, lowest, fitness. The concerned states are the following:
\begin{align}
    {x_{best}}^1 &= \{False, False, False, True, True, True, True, True\} \\
    {x_{best}}^2 &= \{False, True, False, True, True, True, True, True\}
\end{align}
Both of them have fitness ${y_{best}}^1 = {y_{best}}^2 = -23$. We have presented this corner case in order to highlight the role of weights: in situations like the one just shown, assigning a positive weight larger than $1$ to the number of actions is going to make ${x_{best}}^2$ prevail as best state; vice-versa if the residual service matters more.

\subsection{Random graphs}
After showing the capabilities of SAFEX in details, we would like to present an overview analysis on random graphs. We have focused our analysis on sparse random graphs of size 100.\\
We have generated sparse directed graphs using $fast\_gnp\_random\_graph$ function of $NetworkX$. This method, for fixed value of the number of nodes $n$, and of the probability $p$ of creating edges, returns a ${G}_{n,p}$ random graph, also known as an Erd\H{o}s-R\'{e}nyi graph or binomial graph \cite{ErdosRenyi, fast_gnp_random_graphs}.\\
In the ${G}_{n,p}$ model, a graph is constructed by connecting labeled nodes randomly. Each edge is included in the graph with probability $p$, independently from every other edge. In this study we have always chosen $p=1/n$.\\
We have started with sparse graphs of size 100 nodes. We have created different versions of the same random graph, each one with increasing percentage of \switch nodes, from $0.1$ to $0.9$. \source and \user nodes have not been modified in this process, in order to be able to compare the total service before and after the perturbation. The percentage of \switch nodes is relative to the total number of \hub nodes, which have been turned in larger and larger number into switches. It is important to specify that for increasing percentage of switches, the set of \switch nodes contains the ones of smaller percentages as a subset.\\
All the nodes without fathers have been labeled as \source; in the same spirit the ones with no children have been labeled as \user. The rest of the nodes are all \hub nodes initially; a percentage larger and larger is turned into \switch nodes, as stated previously. No node has been labeled as perturbation resistant: the reason for this choice is that we were interested in looking at the propagation of the perturbation in the graph, modulated by the presence of switches. All the edges have weight equal to one. The parameters of the genetic algorithm are the following: $npop$ individuals for the initial population, $ngen$ generations, probability $indpb$ for the attributes to be changed, threshold $tresh$ for applying crossover/mutation. The population to be iterated through the generations is created at first from the initial population, choosing the best $nsel$ selected children, which get modified by mutation or crossover at each step, according to the defined probabilities.
\vspace{1ex}\\
The first graph that we experience is a sparse random graph with only \edgeand logic relationships between nodes. This is the worst case scenario: in fact, in absence of perturbation resistances of any kind, the fault propagation can stop just when arriving to the end of a certain path, after having run it all. In case of \edgeor fathers, instead, just one of the \edgeor fathers have to remain alive in order for the node not to be deleted.\\
We present as first example the perturbation of node $83$ with and without switches. This node is the one with the highest out-degree centrality in the graph. Since the fault propagates from one node to all its children, the breakage of this node is the most disruptive for the graph. Disabling switches, the perturbation of node $83$ causes the break of 17 nodes: $\left[3, 9, 26, 42, 46, 48, 51, 53, 60, 70, 75, 77, 83, 84, 87, 89, 94\right]$. The original service, equal to $27$, turns to be $14$ in the end. In Table \ref{table:83perturbation} we show instead the results for increasing percentage of switches in the graph. The results shown in the table are the best of 10 runs, meaning the ones with the lowest fitness. Moreover, it is worth mentioning that while obtaining these results, we have fixed same weights $w_1 = w_2 = w_3 = 1$ for the three contributions of the fitness. This implies that the genetic algorithm is going to take into account equally the total number of actions and the total number of survived nodes in order to identify the best state, not only the total final service.\\
 What we immediately notice is the fact that, even for the smallest percentage of switches, the number of broken nodes is reduced with respect to their total absence. Moreover, the number of broken nodes diminishes for increasing switch percentage, while the service increases.\\
 Given the set of switches for a fixed percentages, the set for smaller percentages is a subset of the former. Hence, we expect that for increasing percentage of switches the fitness must remain the same or decrease, if increasing the number of \switch nodes a lower minimum has been reached. We notice this behavior in Table \ref{table:83perturbation}.
\begin{table}
    \begin{center}
    \caption{Results for the best switch state for increasing percentage of switches. In the table header, $\# acts$ represents the number of actions, $surv$ the number of survived nodes after the perturbation, $S_{tot}$ is the total final service, while $y_{best}$ is the best fitness of 10 runs. This example refers to the perturbation of node $83$ (out-degree centrality equal to $4$), for a graph of size $100$ nodes. On the left, results for just \edgeand edges are shown, while on the right results for both \edgeor and \edgeand edges are shown. Parameters: $npop = 400$, $ngen = 200$, $indpb = 0.7$, $tresh = 0.4$, $nsel = 100$. }
    \label{table:83perturbation}
    \begin{tabularx}{\textwidth}{X||X|X|X|X||X|X|X|X}
    \toprule
         \multicolumn{9}{c}{\footnotesize{No switches: $83$ survived nodes, $S_{tot} = 14$ (\edgeand); $88$ survived nodes, $S_{tot} = 21$ (\edgeor and \edgeand)}}\\
         \midrule
         \bf \% & \bf \footnotesize{\# acts} & \bf \footnotesize{surv.} & \bf $S_{tot}$ & \bf $y_{best}$ & \bf \footnotesize{\# acts} & \bf \footnotesize{surv.} & \bf $S_{tot}$ & \bf $y_{best}$\\
         \midrule
         \rowcolor{lightgray!30}
         $10$ & $1$ & $84$ & $14$ & $-97$ & $1$ & $90$ & $22$ & $-111$\\
         $20$ & $2$ & $94$ & $22$ & $-114$ & $1$ & $95$ & $26$ & $-120$\\
         \rowcolor{lightgray!30}
         $30$ & $2$ & $94$ & $22$ & $-114$ & $2$ & $96$ & $26$ & $-120$\\
         $40$ & $2$ & $94$ & $22$ & $-114$ & $2$ & $96$ & $26$ & $-120$\\
         \rowcolor{lightgray!30}
         $50$ & $3$ & $96$ & $22$ & $-115$ & $3$ & $98$ & $26$ & $-121$\\
         $60$ & $3$ & $96$ & $22$ & $-115$ & $3$ & $98$ & $26$ & $-121$\\
         \rowcolor{lightgray!30}
         $70$ & $3$ & $96$ & $22$ & $-115$ & $3$ & $98$ & $26$ & $-121$\\
         $80$ & $3$ & $98$ & $23$ & $-118$ & $3$ & $99$ & $26$ & $-122$\\
         \rowcolor{lightgray!30}
         $90$ & $3$ & $98$ & $23$ & $-118$ & $3$ & $99$ & $26$ & $-122$\\
         \bottomrule
    \end{tabularx}
    \end{center}
\end{table}
\vspace{2ex}\newline
The second example that we would like to show is the perturbation of a single node with the smallest out-degree centrality, equal to $1$. A fault departing from this node should have a less severe effect on the graph. Disabling switches, the perturbation of node $93$ causes the break of 4 nodes: $\left[25, 34, 70, 93\right]$. The original service, equal to $27$, turns to be $24$ in the end. In Table \ref{table:93perturbation} we show instead the results for increasing percentage of switches in the graph. We notice that for this perturbation the genetic algorithm converges immediately to a situation in which the only broken node is the one from which the fault starts. This is the best possible result: in fact, no other node is affected. Node $93$ is a \source node, and its only connection is to node $25$, which is a \switch node for all percentages.\\
The lower value of service for $80\%$ of switches can be justified by a smaller number of actions. We again recall that the fitness takes into account the number of actions, the total number of survived nodes and the total final service with the same weight.
\begin{table}
    \begin{center}
    \caption{Results for the best switch state for increasing percentage of switches. In the table header, $\# acts$ represents the number of actions, $surv$ the number of survived nodes after the perturbation, $S_{tot}$ is the total final service, while $y_{best}$ is the best fitness of 10 runs. This example refers to the perturbation of node $93$ (out-degree centrality equal to $1$), for a graph of size $100$ nodes. On the left, results for just \edgeand edges are shown, while on the right results for both \edgeor and \edgeand edges are shown. Parameters: $npop = 400$, $ngen = 200$, $indpb = 0.7$, $tresh = 0.4$, $nsel = 100$.}
    \label{table:93perturbation}
    \begin{tabularx}{\textwidth}{X||X|X|X|X||X|X|X|X}
         \toprule
         \multicolumn{9}{c}{\footnotesize{No switches: $96$ survived nodes, $S_{tot} = 24$ (AND); $97$ survived nodes, $S_{tot} = 24$ (\edgeor and \edgeand)}}\\
         \midrule
         \bf \% & \bf \footnotesize{\# acts} & \bf \footnotesize{surv.} & \bf $S_{tot}$ & \bf $y_{best}$ & \bf \footnotesize{\# acts} & \bf \footnotesize{surv.} & \bf $S_{tot}$ & \bf $y_{best}$\\
         \midrule
         \rowcolor{lightgray!30}
         $10$ & $1$ & $99$ & $25$ & $-123$ & $1$ &$99$ & $25$ & $-123$\\
         $20$ & $1$ & $99$ & $25$ & $-123$ & $1$ &$99$ & $25$ & $-123$\\
         \rowcolor{lightgray!30}
         $30$ & $1$ & $99$ & $25$ & $-123$ & $1$ &$99$ & $25$ & $-123$\\
         $40$ & $1$ & $99$ & $25$ & $-123$ & $1$ &$99$ & $25$ & $-123$\\
         \rowcolor{lightgray!30}
         $50$ & $1$ & $99$ & $25$ & $-123$ & $1$ &$99$ & $25$ & $-123$\\
         $60$ & $2$ & $99$ & $25$ & $-122$ & $1$ &$99$ & $25$ & $-123$\\
         \rowcolor{lightgray!30}
         $70$ & $2$ & $99$ & $25$ & $-122$ & $1$ &$99$ & $25$ & $-123$\\
         $80$ & $6$ & $99$ & $25$ & $-118$ & $1$ &$99$ & $25$ & $-123$\\
         \rowcolor{lightgray!30}
         $90$ & $5$ & $99$ & $25$ & $-119$ & $1$ &$99$ & $25$ & $-123$\\
         \bottomrule
    \end{tabularx}
    \end{center}
\end{table}
\vspace{2ex}\newline
We are now going to present the same perturbation as above, nodes $83$ and $93$, for graphs that present both \edgeor and \edgeand edges. These types of graph have been created from the ones with just \edgeand edges, turning half of the edges to \edgeor edges.\\
The first perturbation studied is as before the one of node $83$, the most connected. In the absence of switches this perturbation causes the breakage of 12 nodes, with a total final service of $21$. As expected, the presence of \edgeor edges mitigates the consequences of the perturbation, even without switches. In Table~\ref{table:83perturbation} we can see that this effect is perpetuated even in the presence of switches. With respect to Table~\ref{table:83perturbation}, the service starts higher even for the smallest percentage of switches, and is always larger comparing the same percentage of switches.\\
Regarding the perturbation of node $93$ instead, in absence of switches three nodes get broken, with a final service of $24$. We can say that the presence of \edgeor edges improves the final state in absence of switches, while in their presence the situation with or without \edgeor edges is more or less the same. However, we have to recall node $93$ presents a favourable situation, being linked by its only edge to a \switch node.\\
We can conclude that in any case the presence of switches improves the final state, both in terms of total number of survived nodes and in terms of total final service.\\
Table~\ref{table:83perturbation_w23eq5} and Table~\ref{table:83perturbation_w1eq5} show the cases in which one of the fitness weights has been alternately set equal to $10$, while the other two remain equal to $1$. Table~\ref{table:83perturbation_w23eq5} exhibits the fact that rewarding the survival of nodes is somehow equivalent to promoting a larger amount of total final service. Table~\ref{table:83perturbation_w1eq5} instead underlines the fact that when the cost of switch flips is large, the need to keep a low number of actions takes place at the expenses of the survival of nodes and of the total final service.

\begin{table}
    \begin{center}
    \caption{Results for the best switch state for increasing percentage of switches. In the table header, $\# acts$ represents the number of actions, $surv$ the number of survived nodes after the perturbation, $S_{tot}$ is the total final service, while $y_{best}$ is the best fitness of 10 runs. This example refers to the perturbation of node $83$ (out-degree centrality equal to $4$), for a graph of size $100$ nodes. On the left, weights $w_1 = w_3 = 1$, $w_2 = 10$, while on the right weights $w_1 = w_2 = 1$, $w_3 = 10$. Parameters: $npop = 400$, $ngen = 200$, $indpb = 0.7$, $tresh = 0.4$, $nsel = 100$.}
    \label{table:83perturbation_w23eq5}
    \begin{tabularx}{\textwidth}{X||X|X|X|X||X|X|X|X}
    \toprule
         \multicolumn{9}{c}{No switches case: $83$ survived nodes, $S_{tot} = 14$}\\
         \midrule
         \bf \% & \bf \footnotesize{\# acts} & \bf \footnotesize{surv.} & \bf $S_{tot}$ & \bf $y_{best}$ & \bf \footnotesize{\# acts} & \bf \footnotesize{surv.} & \bf $S_{tot}$ & \bf $y_{best}$\\
         \midrule
         \rowcolor{lightgray!30}
         $10$ & $1$ & $84$ & $14$ & $-223$ & $1$ & $84$ & $14$ & $-853$\\
         $20$ & $1$ & $91$ & $22$ & $-310$ & $1$ & $91$ & $22$ & $-931$\\
         \rowcolor{lightgray!30}
         $30$ & $2$ & $94$ & $22$ & $-312$ & $2$ & $94$ & $22$ & $-960$\\
         $40$ & $2$ & $94$ & $22$ & $-312$ & $2$ & $94$ & $22$ & $-960$\\
         \rowcolor{lightgray!30}
         $50$ & $3$ & $96$ & $22$ & $-313$ & $3$ & $96$ & $22$ & $-979$\\
         $60$ & $3$ & $96$ & $22$ & $-313$ & $3$ & $96$ & $22$ & $-979$\\
         \rowcolor{lightgray!30}
         $70$ & $3$ & $96$ & $22$ & $-313$ & $3$ & $96$ & $22$ & $-979$\\
         $80$ & $3$ & $98$ & $23$ & $-325$ & $3$ & $98$ & $23$ & $-1000$\\
         \rowcolor{lightgray!30}
         $90$ & $3$ & $98$ & $23$ & $-325$ & $3$ & $98$ & $23$ & $-1000$\\
         \bottomrule
    \end{tabularx}
    \end{center}
\end{table}

\begin{table}
    \begin{center}
    \caption{Results for the best switch state for increasing percentage of switches. In the table header, $\# acts$ represents the number of actions, $surv$ the number of survived nodes after the perturbation, $S_{tot}$ is the total final service, while $y_{best}$ is the best fitness of 10 runs. This example refers to the perturbation of node $83$ (out-degree centrality equal to $4$), for a graph of size $100$ nodes, with weights $w_2 = w_3 = 1$, $w_1 = 10$. Parameters: $npop = 400$, $ngen = 200$, $indpb = 0.7$, $tresh = 0.4$, $nsel = 100$.}
    \label{table:83perturbation_w1eq5}

    \begin{tabularx}{\textwidth}{X||X|X|X|X}
    \toprule
         \multicolumn{5}{c}{No switches case: $83$ survived nodes, $S_{tot} = 14$}\\
         \midrule
         \bf \% & \bf \footnotesize{\# acts} & \bf \footnotesize{surv.} & \bf $S_{tot}$ & \bf $y_{best}$\\
         \midrule
         \rowcolor{lightgray!30}
         $10$ & $0$ & $83$ & $14$ & $-97$\\
         $20$ & $1$ & $91$ & $22$ & $-103$\\
         \rowcolor{lightgray!30}
         $30$ & $1$ & $91$ & $22$ & $-103$\\
         $40$ & $1$ & $91$ & $22$ & $-103$\\
         \rowcolor{lightgray!30}
         $50$ & $1$ & $91$ & $22$ & $-103$\\
         $60$ & $1$ & $91$ & $22$ & $-103$\\
         \rowcolor{lightgray!30}
         $70$ & $1$ & $91$ & $22$ & $-103$\\
         $80$ & $1$ & $91$ & $22$ & $-103$\\
         \rowcolor{lightgray!30}
         $90$ & $1$ & $91$ & $22$ & $-103$\\
         \bottomrule
    \end{tabularx}
    \end{center}
\end{table}

\section{Conclusions}
\label{sec:conclusions}
In this work, we have presented a method able to perform risk analysis, together with the consequent fault system diagnostics, on complex interconnected systems. Everything that has been presented in this paper is implemented in GRAPE software \cite{GRAPE}.\\
Weighted graphs describe preferable itineraries for the flow of commodities, with respect to areas in which the course can be less favoured; everything is efficiently integrated in the graph paradigm. Our focus in this dissertation has been devoted to the introduction of the new concept of \textit{service}. Its creation re-defines the different roles of the nodes within the network, and allows a quantification of its residual operational capacity. In particular, its value is preserved and maximized by an automatic identification of the best configuration of the set of \switch nodes, that are involved in limiting the cascade effect within GRAPE simulation of a fault propagation in the system.\\
The different types of perturbations available in the software are explained, and several examples are given on model graphs. Results obtained on random sparse graphs are also shown, highlighting how fault limitation is improved by the presence of an increasing number of switches.
The methodology illustrated here is of immediate application on industrial plants systems of different nature. The graph standard is general enough to be applied to diverse cases, integrated with a precise estimate of the resources still available in the system. For these reasons, this software can be employed in the development of a digital twin for fault analysis, from the early stages. The scalability of the graph concept towards larger system is of immediate transfer, and is going to be analyzed in even further details in the future. In fact, future developments are going to include querying the software with realistic plants of increasing complexity. This implies the extension of GRAPE to graphs of increasing size.\\
Finally, another possible extension of this work can be the coupling of graph topology with realistic physical quantities on nodes or edges; it can be of relative easy implementation within our software, and could definitely be of wide use within realistic industrial applications.

\section*{Acknowledgements}
This work was partially supported by European Union Funding for Research and Innovation ---
Horizon 2020 Program --- in the framework of European Research Council
Executive Agency: H2020 ERC CoG 2015 AROMA-CFD project 681447 ``Advanced
Reduced Order Methods with Applications in Computational Fluid Dynamics'' P.I.
Gianluigi Rozza.

\bibliographystyle{abbrv}
\bibliography{biblio}

\end{document}